\documentclass[aps,twocolumn, prb,showpacs,groupedaddress]{revtex4}
\usepackage{graphicx}

\begin{document}

\title{Correlation energy in a spin polarized two dimensional electron liquid in the high density limit}
\author{Stefano Chesi}
\author{Gabriele F. Giuliani}
\affiliation{Department of Physics, Purdue University,
West Lafayette, IN 47907, USA}

\date{\today}

\begin{abstract}
We have obtained an analytic expression for the ring diagrams contribution to
the correlation energy of a two dimensional electron liquid as a function of the uniform
fractional spin polarization. Our results can be used to improve on the interpolation
formulas which represent the basic ingredient for the construction of modern
spin-density functionals in two dimensions.
\end{abstract}


\maketitle

\setlength\arraycolsep{0pt}

The electron liquid, an ideal model in which electrons interact via a Coulomb potential
in the presence of a uniform rigid neutralizing background, represents a fundamental paradigm 
for the understanding of condensed matter systems.\cite{TheBook} Although for homogeneous 
states the ground state properties are completely determined by the number density $n$, it can
be useful and often necessary to study the properties of the system in the presence of a
given uniform spin polarization. Accordingly, for the case under consideration of a 
two dimensional system, alongside the traditionally employed dimensionless density parameter 
$r_s = \sqrt{\pi n a_B^2}$ we will consider the effects of polarization as parameterized by 
$p=\frac{2 S_z}{\hbar n}$, where $S_z$ is the uniform spin polarization density.

While in the general uniform case the total energy of the system can be accurately 
obtained only by numerical means and is available in the form of interpolation formulas 
of Monte Carlo results,\cite{attaccalite02,TheBook} an exact analytic treatment is possible in 
the high density limit in the form of a perturbative expansion in $r_s$ which in the 
case is vanishingly small.\cite{comment_noSDW_at_high_dens}
In this limit the first-order correction to the noninteracting energy simply coincides 
with the familiar exchange energy, which has a simple analytic formula proportional
to $r_s^{-1}$. The remaining correction is referred to as the correlation energy.\cite{TheBook} 

The high-density expansion of the correlation energy was studied in the 
classic Ref.~\onlinecite{rajagopal77}. The final expression for the paramagnetic case 
(in $Ry$ units) reads:
\begin{equation}\label{correl}
\mathcal{E}_c(r_s,p=0)= -0.385  -\frac{2\sqrt{2}}{3\pi} (10-3\pi)\, r_s \ln r_s + \ldots  ~,
\end{equation}
while the $p=1$ result can be also obtained, by making use of a simple 
transformation.\cite{rajagopal77}

For generic values of the polarization, the coefficients of the expansion (\ref{correl})
become functions of $p$.

The constant term in (\ref{correl}) is obtained from second-order perturbation 
theory and is the sum of two distinct contributions. The first is the second-order 
exchange energy, which is independent of $p$ and can be calculated 
analytically.\cite{isihara80} 
The second one stems from the second-order direct energy term, commonly referred 
to as the first ring diagram. The latter has been recently evaluated numerically 
for generic values of $p$ in Ref.~\onlinecite{seidl04}. These contributions are 
accurately represented by the interpolation formula of Ref.~\onlinecite{attaccalite02}.

The next perturbative terms are in general divergent and a finite result is obtained 
upon exact summation of the infinite series of the most diverging contributions, the 
remaining ring diagrams. This leads to the sub-leading $r_s \ln r_s$ term.

This elegant method was originally developed for the corresponding unpolarized
three dimensional case,\cite{gellmann57} a problem in which the ring diagrams sum up to give
the leading $\ln r_s$ contribution to the correlation energy.\cite{comment_3Dvs2D_rings}
For the three dimensional case the exact polarization dependence was also 
determined.\cite{wang91} 

The generic dependence of the $r_s \ln r_s$ term on $p$ is obtained in this work.
The complete formula for the presently known leading terms of $\mathcal{E}_c(r_s,p)$ 
is provided for reference in the Appendix.

%The paper is organized as follows:
%
%\section{Formulation}

Following Ref.~\onlinecite{rajagopal77}, the value of the generic diverging ring 
diagram of order $n$ is obtained from the expression:
\begin{equation}\label{rings_def}
f^{(n)}_R(p)=
-\frac{(-1)^n}{\pi n r_s^2}
\int_{-\infty}^{+\infty} {\rm d}u
\int_0^\infty q^2 {\rm d}q
\left(
\frac{Q_q(u)r_s}{2\sqrt{2}\pi q}
\right)^n ,
\end{equation}
where $n$ is a positive integer larger than 1. 

The explicit expression for $Q_q(u)$ is given by:
\begin{equation} \label{Qdef}
Q_q(u)
= \sum_{\sigma=\uparrow,\downarrow} 
\int 
\frac{
{\rm d} {\bf k}}{q} 
\, 
\frac{
q+2 k_x
}{
(\frac{q}{2}+ k_x)^2+ u^2
} 
\,
n_\sigma(k)(1-n_\sigma(k')) ~, \nonumber
\end{equation}
where $k'=\sqrt{(k_x+q)^2+k_y^2}$. The polarization dependence of $Q_q(u)$ 
is implicitly determined by the occupation functions:
\begin{equation}\label{occup}
n_\sigma(k)=\theta(k_{\sigma}-k) ~.
\end{equation}
For each of the two spin orientations, the Fermi wave vector is obtained
from the relation:
\begin{equation}\label{fermiwv}
k_{\uparrow (\downarrow)}~=~ \sqrt{1\pm p}  ~,
\end{equation}
where for convenience we have rescaled the wave vectors by $k_F=\sqrt{2\pi n}$. 

The expression corresponding to the generic ring diagram of Eq.~(\ref{rings_def}) has 
a (formal) dependence of $r_s^{n-2}$, but in reality for $n \geq 3$ it involves a 
diverging integral. 
The sum to infinite order is:
\begin{eqnarray}\label{rings_sum}
f_R(p)=\frac{1}{\pi r_s^2}
\int_{-\infty}^{+\infty} {\rm d}u
\int_0^\infty q^2 {\rm d}q
\Bigg[
\ln \left(
1+\frac{Q_q(u)r_s}{2\sqrt{2}\pi q}
\right)
\quad\\
-\frac{Q_q(u)r_s}{2\sqrt{2}\pi q} 
+\frac12 
\left( 
\frac{Q_q(u)r_s}{2\sqrt{2}\pi q}
\right)^2
\Bigg]~,  \nonumber 
\end{eqnarray}
which is instead a converging integral whose leading contribution in the $r_s\to 0$ limit 
is again proportional to $r_s \ln r_s$ and is determined by the small-$q$ 
integration region.\cite{comment_no_2_ring_constant} 
Accordingly it is sufficient to employ here the limiting value $Q_{q=0}(u)$, 
as obtained from (\ref{Qdef}). The result can be written as:
\begin{eqnarray} \label{Qresult}
Q_{0}(u) =
2\pi 
[ \, R(u/k_\uparrow) +
R(u/k_\downarrow) \, ]
~, \qquad
\end{eqnarray}
where the function $R(u)$ is defined as:
\begin{equation}\label{Rdef}
R(u)=1-\frac{1}{\sqrt{1+1/u^2}} ~.
\end{equation}

To this point our discussion follows Ref.~\onlinecite{rajagopal77}, where the implicit expression
(\ref{rings_sum}) as well as Eq.~(\ref{Qresult}) were originally provided for a generic value of $p$. There however, the explicit evaluation was only done for $p=0$. 
%the $p=1$ case being obtained through
%an elegant scaling formula. 
In the general case, using (\ref{Qresult}) and performing 
in (\ref{rings_sum}) the wave vector integration (up to an arbitrary upper limit), one 
extracts the leading contribution in $r_s$:
\begin{equation}\label{ring_approx}
f_R(p)\simeq -\frac{r_s \ln r_s}{ 3\sqrt{2}(2\pi)^4} 
\int_{-\infty}^{+\infty} [Q_0(u)]^3 \, {\rm d}u  ~,
\end{equation}
which gives the standard result (\ref{correl}) by making use of $Q_0(u)=4\pi R(u)$,
i.e.
\begin{equation}\label{ring_approx_p_equal0}
f_R(0) ~\simeq~ -\frac{2\sqrt{2}}{3\pi} (10-3\pi)\, r_s \ln r_s ~.
\end{equation}

We find that the integral in (\ref{ring_approx}) can be performed exactly in the general case. 
The result can be expressed in terms of the function
\begin{equation}\label{Fdef}
F(x,y)=\int_{-\infty}^{+\infty} 
[R(u/x)]^2 R(u/y) \, {\rm d}u ~,
\end{equation}
which has the explicit expression:
%\cite{Mathematica5.2}
\begin{equation}\label{Fresult}
F(x,y)=4(x+y)-\pi x
-4 x E(1-\frac{y^2}{x^2})
+\frac{2 x^2 \arccos\frac{y}{x}}{\sqrt{x^2-y^2}}~.
\end{equation}
Here $E(x)$ is the complete elliptic integral of the second type,\cite{Abramowitz65} 
and one should use the identity 
$\frac{\arccos\frac{y}{x}}{\sqrt{x^2-y^2}}= 
\frac{{\rm arccosh}\frac{y}{x}}{\sqrt{y^2-x^2}}$ 
for $y>x$. At $p=0$ one only needs the value $F(1,1)=10-3\pi$.

The result for the ring diagrams at finite polarization can be compactly and elegantly expressed in terms of the 
corresponding polarization scaling function $I_R(p)$, defined as:
\begin{equation}\label{defI}
I_R(p) ~= ~ \lim_{r_s \to 0} \frac{f_R(p)}{f_R(0)}  ~.
\end{equation}
The final expression is given by:
\begin{equation}\label{ring_result}
I_R(p) = \frac{1}{8} \,
\left(
k_\uparrow+k_\downarrow
+3 \, \frac{
F(k_\uparrow,k_\downarrow)
+ F(k_\downarrow,k_\uparrow)
}{10-3\pi} \right) ~, \nonumber
\end{equation}
which readily gives the correct value at $p=0$. At $p=1$, using $F(\sqrt{2},0)=F(0,\sqrt{2})=0$,
we obtain the known result\cite{rajagopal77} $I_R(1)=\frac{\sqrt{2}}{8}$.

%%%%%%%%%%%%%%%%%%%%%%%%%%%%%%%%%%%%%%%%%%%%%%%%%%%%%%%%%%%%%%%%%%%%%%%%%%%%%%%%%%%%%%%%%%%%%%%%%%%%%
\begin{figure}
\begin{center}
\includegraphics[width=0.43\textwidth]{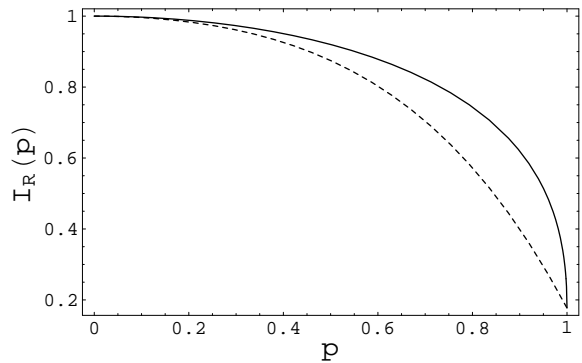}
\caption{\label{logcoeff} Plot of the scaling factor $I_R(p)$ (defined in Eq.~(\ref{defI})) 
as a function of the fractional polarization $p$. Solid line: exact expression
(\ref{ring_result}). Dashed line: Eq.~(\ref{ringsMC}) as obtained from
the interpolation formula proposed in Ref.~\onlinecite{attaccalite02}.}
\end{center}
\end{figure}
%%%%%%%%%%%%%%%%%%%%%%%%%%%%%%%%%%%%%%%%%%%%%%%%%%%%%%%%%%%%%%%%%%%%%%%%%%%%%%%%%%%%%

As it turns out the exact result $I_R(p)$ is not well reflected in the most recent 
interpolation formulas of Monte-Carlo calculations provided in the literature. In particular, 
from the correlation energy formula of Ref.~\onlinecite{attaccalite02}, denoted 
here as $\mathcal{E}_c^{MC}(r_s,p)$, the following limit is obtained:
\begin{eqnarray}\label{ringsMC}
I_R^{MC}(p) &&=
\lim_{r_s \to 0} \frac{\mathcal{E}_c^{MC}(r_s,p)-\mathcal{E}_c^{MC}(0,p)}{f_R(0)} 
\nonumber\\
&&\simeq  ~ 1- 0.3932 \, p^2 -  0.4297 \, p^4  ~, 
\end{eqnarray}
which is compared in Figure \ref{logcoeff} to the exact result of Eq.~(\ref{ring_result}).
The difference is remarkable, even if the specific aim of Ref.~\onlinecite{attaccalite02} 
is to address the polarization dependence of the whole correlation energy.
Agreement of $I_R^{MC}(p)$ with the exact result is only achieved for $p=0$ and
$p=1$, values known from the extant literature.
A noticeable failure is the behavior near $p=1$, where the polynomial (\ref{ringsMC}) 
gives a finite slope while the leading term in the exact expression is:
\begin{equation}
I_R(p)\simeq \frac{\sqrt{2}}{8}+\frac{14-3\pi}{4(10-3\pi)} \, \sqrt{1-p} ~.
\end{equation}
Around $p=0$, while being correctly quadratic in $p$, Eq.~(\ref{ringsMC}) displays an 
incorrect coefficient. The exact coefficient is given by:
\begin{equation}\label{ring_small}
I_R(p) \simeq 1 - \frac{168 - 45 \pi}{160 (10 - 3 \pi)}  \, p^2 
\simeq 1- 0.2893 \, p^2  ~. 
\end{equation}

The disagreement between $I(p)$ and the Monte Carlo based interpolation
$I^{MC}(p)$ is hardly surprising for the latter was obtained by sampling the energy
at a number of polarization values for each of the values $r_s=1,2,5,10$, which are clearly
outside of the $r_s \ll 1$ perturbative regime.\cite{commento_RPAsmallrs}
In practice, $I(p)$ refers to the sub-leading term in the density expansion, 
so that it gives only small corrections to the total energy. Nevertheless, incorporating 
this exact formula would certainly result in an improved empirical expression for the 
polarization dependence of the correlation energy.

We conclude by noting that a similar, yet more involved, calculation can be carried out 
for the high density limit of the electron liquid correlation energy in the presence 
of Rashba spin orbit. An analysis of this interesting and timely problem is provided 
elsewhere.\cite{unpub_ChesiGFG}

\appendix*
\section{}

For ease of reference, we collect here the explicit form of all the leading terms 
contributing to the perturbative expansion of the total energy of the two dimensional 
electron liquid at finite polarization. The general formula (in Ry units) reads:
\begin{equation}\label{totalEn}
\mathcal{E}(r_s,p)=\mathcal{E}_K(r_s,p)+\mathcal{E}_x(r_s,p)+\mathcal{E}_c(r_s,p) ~.
\end{equation}
In this expression $\mathcal{E}_K(r_s,p)$ represents the noninteracting kinetic 
energy and is given by:
\begin{equation}\label{Ex}
\mathcal{E}_K(r_s,p)=\frac{1+p^2}{r_s^2} ~,
\end{equation}
while $\mathcal{E}_x(r_s,p)$ represents the exchange energy:
\begin{equation}\label{Ekin}
\mathcal{E}_x(r_s,p)=-\frac{8\sqrt{2}}{3\pi} \,\frac{(1+p)^{3/2}+(1-p)^{3/2}}{2 r_s} ~.
\end{equation}
The expansion for the correlation energy takes the form:
\begin{equation}\label{Ec}
\mathcal{E}_c(r_s,p) =\mathcal{E}_2(p)-\frac{2\sqrt{2}}{3\pi}(10-3\pi) \, 
I_R(p) \,r_s \ln r_s  + \ldots~,
\end{equation}
where the omitted corrections are of order ${\cal O}(r_s)$. 
In this expression $I_R(p)$ is defined by Eqs.~(\ref{ring_result}), (\ref{Fresult}) 
and (\ref{fermiwv}). The density independent term is in turn given by:
\begin{equation}\label{E2}
\mathcal{E}_2(p) =\mathcal{E}_2^{(b)}+\mathcal{E}_2^{(r)}(p)~,
\end{equation}
where:\cite{isihara80}
\begin{equation}\label{E2b}
\mathcal{E}_2^{(b)} = 0.2287 ~,
\end{equation}
and\cite{seidl04}
\begin{equation}\label{E2r}
\mathcal{E}_2^{(r)}(p) = - 0.6137 \, I_2(p) ~,
\end{equation}
where the scaling function $I_2(p)$ is given by:
\begin{equation}\label{I2}
I_2(p) =1-\frac{(1+p)\ln(1+p)+(1-p)\ln(1-p)}{4 \ln 2}-\frac{\delta f(p)}{2} ~,
\end{equation}
with\cite{comment_Seidl04dip}
\begin{equation}\label{df}
\delta f(p) \simeq 0.0636 \, p^2-0.1024 \, p^4 +0.0389 \, p^6 ~.
\end{equation}

%%%%%%%%%%%%%%%%%%%%%%%%%%%%%%%%%%%%%%%%%%%%%%%%%%%%%%%%%%%%%%%%%%%%%%%%%%%%%%%%%%%


\begin{thebibliography}{17}
\bibitem[{com({\natexlab{c}})}]{comment_no_2_ring_constant}
\bibinfo{note}{Notice that by construction the density independent contribution
  from the $n=2$ ring diagram has been omitted. This is the quantity studied in
  Ref.~\onlinecite{seidl04}.}
\bibitem[{com({\natexlab{d}})}]{commento_RPAsmallrs}
\bibinfo{note}{It should be kept in mind that in practice the perturbative
  theoretic expression of the energy of the electron liquid based on the ring
  diagrams does provide a quantitavely accurate description only at very small
  values of $r_s$ (see also Ref.~\onlinecite{TheBook}).}
\bibitem[{com({\natexlab{e}})}]{comment_Seidl04dip}
\bibinfo{note}{It should be noted that as in the original paper, Eq.~(\ref{df})
  does not account for a small dip near $p =1$ seemingly uncovered in the
  numerical work of Ref.~\onlinecite{seidl04}.}
\end{thebibliography}
\end{document}